\documentclass[aps,pra,twocolumn, portrait, preprintnumbers]{revtex4}

\usepackage{placeins}
\usepackage{makecell}
\usepackage{url}

\usepackage{graphicx}
\usepackage{dcolumn}
\usepackage{bm}
\usepackage{url}
\usepackage{xcolor}
\usepackage{float}
\usepackage[pdftex,colorlinks=true,
    linkcolor=blue,
    filecolor=blue,
    urlcolor=blue,
    citecolor=blue,
    plainpages=false,
    bookmarksopen=true]{hyperref} 

\begin{document}

\title{Assessing the Needs of the Quantum Industry}

\author{Ciaran Hughes}
\affiliation{Fermi National Accelerator Laboratory, Batavia, IL, 60510, USA}

\author{Doug Finke} 
\affiliation{Quantum Computing Report, Tustin, CA, 92780, USA}

\author{Dan-Adrian German} 
\affiliation{Luddy School of Informatics, Indiana University,  Bloomington, IN 47408, USA}
\affiliation{Quantum Science and Engineering Center, Indiana University, Bloomington, IN 47408, USA}

\author{Celia Merzbacher} 
\affiliation{SRI International, 1100 Wilson Blvd, Suite 2800, Arlington VA 22209 USA}

\author{Patrick M. Vora} 
\affiliation{Department of Physics and Astronomy, George Mason University, Fairfax, VA 22030 USA}
\affiliation{Quantum Science and Engineering Center, George Mason University, Fairfax, VA 22030 USA}

\author{H. J. Lewandowski}
\affiliation{JILA, National Institute of Standards and Technology and the University of Colorado, Boulder, CO 80309, USA}
\affiliation{Department of Physics, University of Colorado Boulder, Boulder, CO 80309, USA}

\preprint{ FERMILAB-PUB-21-381-T}

\begin{abstract}
Quantum information science and technology (QIST) has progressed significantly in the last decade, such that it is no longer solely in the domain of research labs, but is now beginning to be developed for, and applied in, industrial applications and products. With the emergence of this new quantum industry, a new workforce trained in QIST skills and knowledge is needed. To help support education and training of this workforce, universities and colleges require knowledge of the type of jobs available for their students and what skills and degrees are most relevant for those new jobs. Additionally, students need to know how to tailor their degrees to best align with the current needs of the quantum industry. We report on the results from a survey of 57 companies in the quantum industry, with the goal of elucidating the jobs, skills, and degrees that are relevant for this new workforce. We find a range of job opportunities from highly specific jobs, such as quantum algorithm developer and error correction scientist, to broader jobs categories within the business, software, and hardware sectors. These broader jobs require a range of skills, most of which are not quantum related. Further, except for the highly specific jobs, companies that responded to the survey are looking for a range of degree levels to fill these new positions, from bachelors to masters to PhDs. With this knowledge, students, instructors, and university administrators can make informed decisions about how to address the challenge of increasing the future quantum workforce. 

\end{abstract}

\maketitle

The last few years have seen a rapid growth in progress towards the development of commercially available quantum-enabled products. The most well-known highlight was a demonstration of the quantum advantage, where a 53-qubit quantum computer achieved a benchmark not possible within a reasonable timescale on a classical computer \cite{arute2019}. A similar result was reported soon after \cite{pan2020}. Other commercial systems have also been demonstrated, including a trapped-ion quantum computer with a charge-couple device architecture \cite{pino2021}, and an 11-qubit quantum computer using trapped ions \cite{wright2019benchmarking}.  In parallel with these new achievements, there has been a significant investment into quantum information science and technology (QIST) from the U.S. federal government, enabled by the National Quantum Initiative Act \cite{NQI}. This investment has funded three National Science Foundation Quantum Leap Challenge Institutes \cite{qlci}, as well as five Department of Energy National Quantum Information Science Research Centers \cite{doe}. These centers include collaborations among multiple universities, industrial partners, and Federal labs, with the aim of creating a healthy ecosystem to advance QIST in the U.S. and promote the use of QIST for national security, economic growth, and scientific leadership. 

Emerging from these new scientific breakthroughs and with the influx of resources is a new ``quantum industry'', which we define as including all companies that use QIST in their business/products or provide technologies that enable such business/products. To facilitate this growing industry, the National Institute of Standards and Technology established the the Quantum Economic Development Consortium (QED-C) in 2018. This organization was designed to ``enable and grow a robust commercial quantum-based industry and associated supply chain in the United States.'' \cite{qedc} With this developing industry comes a demand for a new quantum-literate and quantum-expert workforce. To meet the anticipated need of this growing workforce, several programs have begun, including, for example, National Q-12 Education Partnership (Q2Work), which is a ``consortium that will expand access to K-12 quantum learning tools and inspire the next generation of quantum leaders,'' \cite{q2work}, accessible high school resources \cite{Hughes2021, Hughes:2020t}, enhanced undergraduate degree programs \cite{asfaw2021building}, new master's level degree programs \cite{Aiello_2021}, workshops \cite{duke}, and industrial education programs \cite{ibm,azure}.

For all of these quantum education and workforce development efforts to be successful, there has to be knowledge about the characteristics of this new quantum workforce (i.e., what types of jobs are available and what degrees and skills are needed for these jobs?). There has been one research study on this topic, which used in-depth interviews with 21 companies from the quantum industry to understand the types of jobs available, the skills and knowledge required for those jobs, as well as the current career pathways into those positions. \cite{fox2020} They also reported on hiring challenges and possible ways higher-education institutions could address those challenges. This was a good first step towards informing students, educators, and policy makers about steps they could each take to benefit from, and contribute to, the quantum workforce. However, the number of companies participating in that study was limited due to the qualitative nature of the interviews. A broader quantitative study is also needed to understand how generalizable these initial results might be, and to what extent additional skills, knowledge, and degrees are needed for the quantum industry across the U.S.  

Here, we present the results from such a study of a broader selection of quantum companies. The goal of this work is to inform all stakeholders (i.e., students, university educators and administrators, policy makers, funding agencies, and quantum companies) about the needs of the quantum industry. In doing so, we answer three main research questions: What types of jobs are available?; Which skills are important for those jobs?; and Which degrees are important for those jobs? The answers to these questions will help guide the diverse quantum education and workforce development programs, so they are able to reach their goals for creating a vibrant quantum workforce. Without data driven measures used to inform the various stakeholders, the success of such programs is unclear. 

\section*{Previous investigation of the Quantum Workforce}

The need for an expanded quantum workforce has closely tracked the recent growth of the quantum industry. Therefore, there has only been a single previous systematic study of needs of the quantum industry. 
As mentioned above, Fox et al. interviewed 26 people (at 21 companies) who hire and supervise new employees in quantum-related jobs \cite{fox2020}. The results of that study included information on the common types of jobs, the valued skills and knowledge (both broadly over many jobs and also for each individual job role), how these skills and knowledge were gained by their current employees, and what employers viewed as the most significant hiring challenges. 

The authors found that most companies had job roles categorised as ``engineer'' (95\% of companies) and ``experimental physicist'' (86\%), and about half had employees working as ``theorists'' (57\%) and ``technicians'' (43\%). Almost all companies generally valued classical programming skills (90\% of companies), using statistical methods for data analysis (90\%), and laboratory experience (87\%). With regard to how these skills were obtained, companies reported the degrees held by their employees, including a Ph.D. in physics (95\% of companies) and a B.S. Engineering (57\%). 
The companies also reported that they expected their employees to learn the necessary skills ``on the job'' (95\% of companies interviewed), either through independent learning or from coworkers. Overall, there was no consensus about the most critical hiring challenges, about one-third of companies stated that hiring quantum information theorists and those with analog electronics skills was a challenge. 

We build on this prior work and report on the more specific job types and skills, knowledge, and degrees associated with those individual jobs. With our quantitative study methodology, we are also able to look at correlations between different skills and knowledge to motivate development of education programs and guide students in their educational choices.

\begin{figure*}[tbhp]
\centering
\includegraphics[width=.9\textwidth]{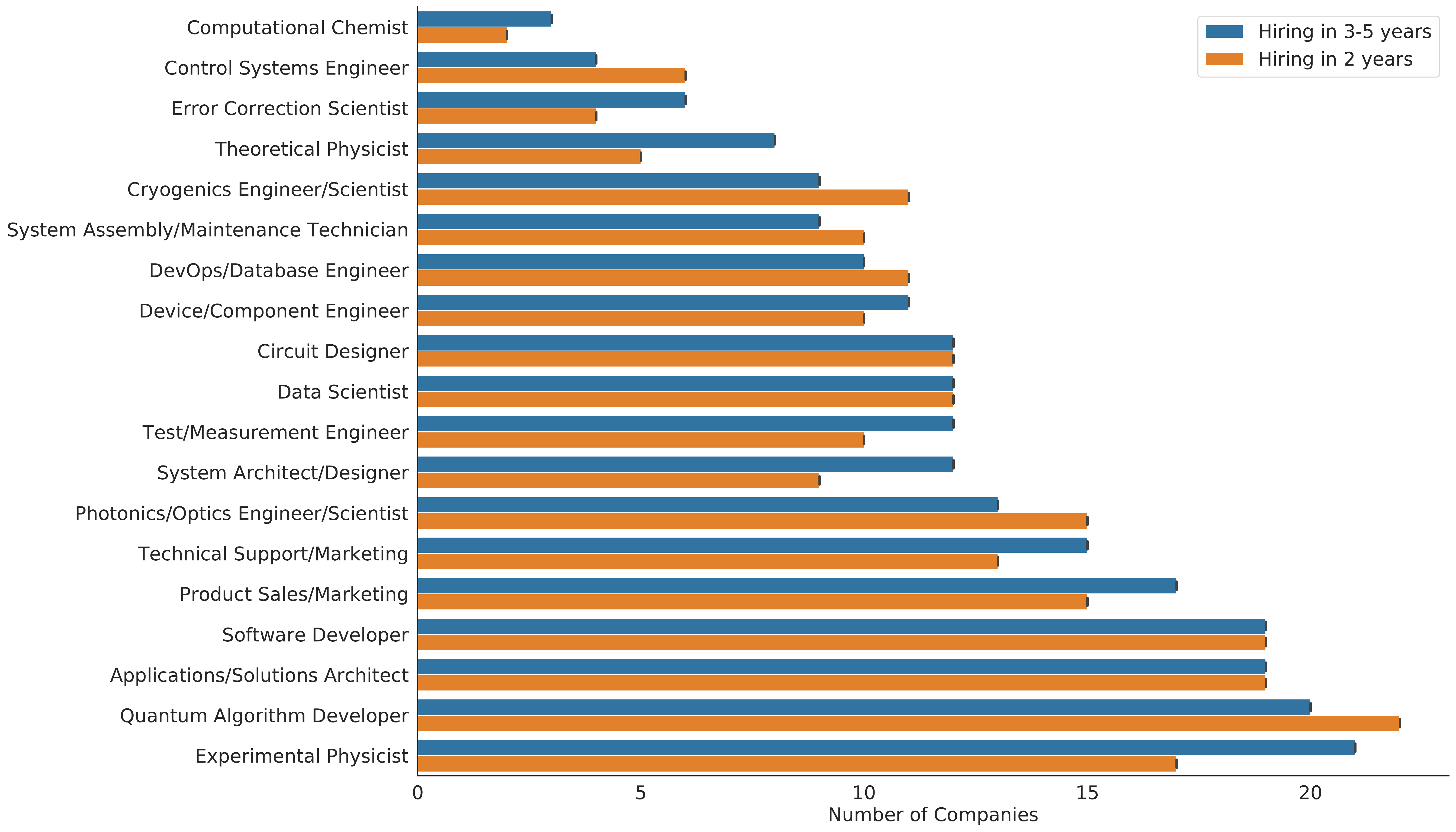}
\caption{The number of companies out of 57 (x-axis) that indicated they would hire a job type (y-axis) over the next 2 years (blue) or the next 3-5 years (orange). Note that the data are similar over the two time periods, as discussed in the text.}
\label{fig:commonjob}
\end{figure*}

\section*{Methodology}

Members of the QED-C Workforce Technical Advisory Committee developed a survey in the summer and fall of 2020 to gather information from QED-C member companies about their workforce needs. The goal of the survey was not to collect projected numbers of how many new hires the companies planned to make, but rather the quantum related jobs roles and the degrees and skills needed for those jobs. This decision was made to encourage companies to participate, as hiring forecasts are seen as propriety information in such a competitive industry. Most of the survey was in a closed-response format, with two open-response questions at the end. Here, we consider only the closed-response questions. The choices for the closed-response questions were chosen by examining over 400 quantum-related jobs ads that were posted online during the summer of 2020, and compiling a list of most common terms used in job titles and required skills. These were condensed into a set of the most common items (19 for job roles and 31 for skills, with an text box for ``other'') to serve as choices for the survey. (See Supplementary Information for the complete survey.)

The structure of the survey began by asking ``What is the primary nature of your company’s business? (Select at most three answers)'' and then went on to ask ``What directly quantum-related work roles will your company need within the next 2 years? and next 3\textendash5 years?'' For each job selected, two questions appeared, which were ``What primary skills and knowledge will be needed for the [job selected] role?'' and ``What is the preferred education or experience for [job selected] for new hires?'' The last question on the survey asked ``Please include a name and contact information if you are willing to be contacted if there are follow up questions.'' Thus, the survey was anonymous for all unless they answered this last question. In total, 131 companies received a request to complete the survey, with 57 choosing to do so. Of the 57 respondents, 27 filled in the last question identifying their company.  

The main limitation of this study is the unknown bias in who decided to complete the survey. We do not know if the 57 companies are a representative sample of the QED-C companies asked to complete the survey, or if QED-C member companies are representative of the quantum industry in the US generally. (See Supplemental Information about how we tried to understand this possible bias and a plot of the self-reported nature of the companies' business.)

As there is a broad range of companies represented in our data, we present only the ``important'' skills or degrees, so as to focus on items that are common across the quantum industry as a whole $-$ this information will be more useful for understanding the general industry, rather than one sub-sector. We define an important skill or degree as one in which more than $50\%$ of respondents to that part of the survey said that a skill or degree was needed for a particular job. That is, we define importance by consensus.


\begin{figure*}[]
\centering
\includegraphics[width=.85\textwidth]{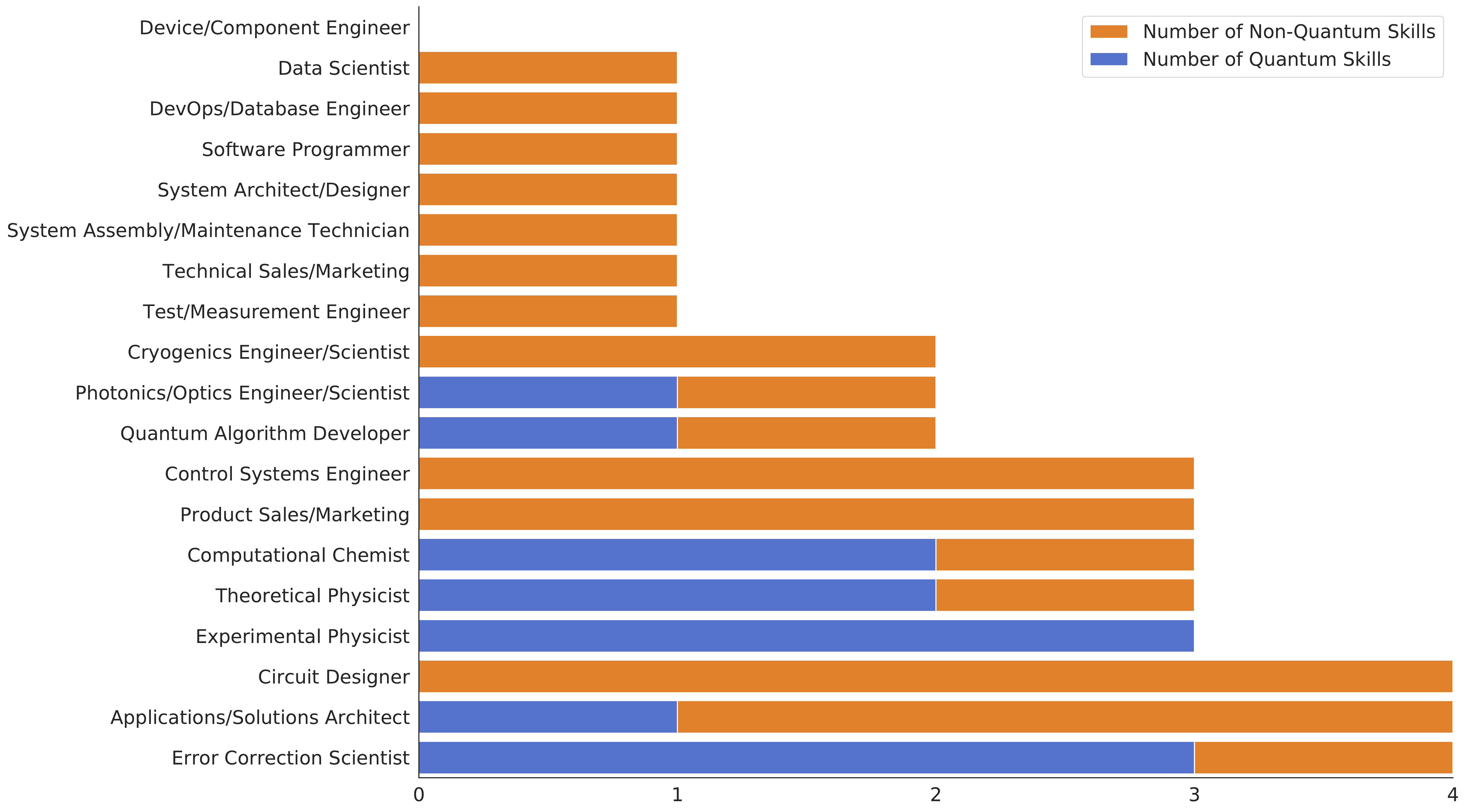}
\caption{The number of important quantum (blue) and non-quantum (orange) skills needed for each job type. Note that not all job types in the quantum workforce need important quantum skills, as discussed in the text.  }
\label{fig:importantskills}
\end{figure*}

\section*{What types of jobs are available?}
It is important for all stakeholders to understand the range of job roles that are expected within the next five years. Such information will allow students to focus their educational choices and for universities to direct their resources at the highest impact areas of skill development. Figure \ref{fig:commonjob} shows the number of companies who said they would be hiring in a particular job role in the next few years. An important observation from this graph is that the number of companies who will be hiring in any one job role over the next two years and the next three to five years is nearly the same. Thus, the distribution in the $\emph{types}$ of jobs is not predicted to change in the near term. This is not to say that the number of open positions for a particular job will be the same over this time-span, but rather the nature of the jobs at these companies is not predicted to change. This is in line with the idea that although there has been great progress in the development of quantum technologies, the fundamental nature of the industry is not likely to change dramatically in the next five years. Currently, there is still a considerable amount of research and development and significantly less production of products. This state of the industry will likely continue in the next few years. 

It is also important to note that these data do not say what will be the most common jobs available, as we did not ask how many of each type of position the various companies will be looking to fill. Thus, although the highly specialized jobs (e.g., error correction scientist) will be present in almost half of the companies surveyed, it could be that each of those companies may be looking to hire only a few in such a role. In contrast, companies that wish to hire, for example, a test and measurement engineer, may each want to hire to fill a considerable number of those positions. 

These data do allow those developing new degree programs to be confident that, as least in the near term, the range of jobs will not dramatically decrease or increase. Thus, programs that aim to help students obtain the jobs listed will be relevant for some time to come.

\begin{figure*}[tbhp]
\centering
\includegraphics[width=.9\textwidth]{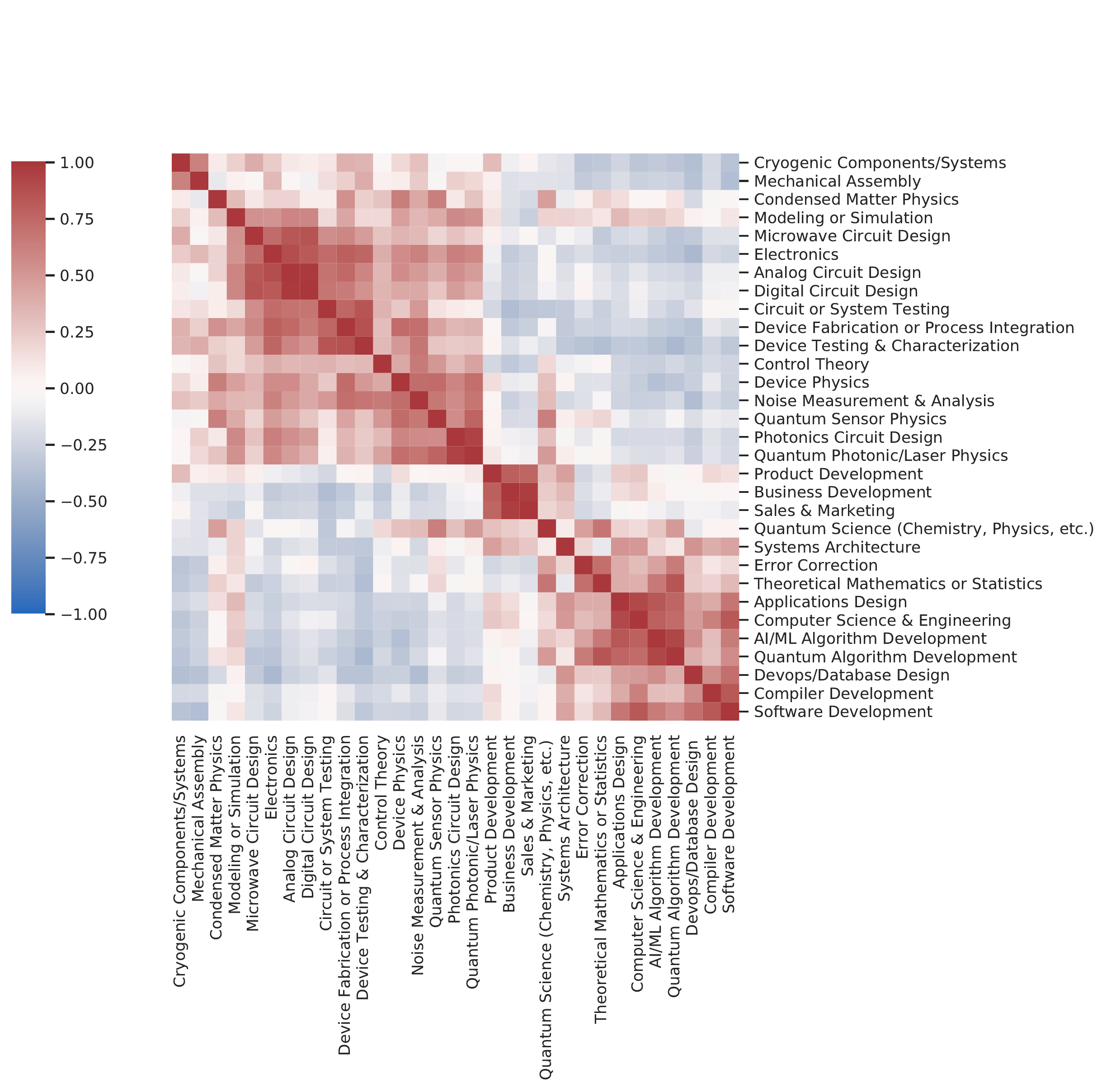}
\caption{Survey respondents listed which skills were needed (or not) for each job. From these data, we compute the (Pearson) correlation between skills, by averaging between job types. The three diagonal blocks can be characterised by hardware, business, and software, indicating that skills are siloed within these three sectors. }
\label{fig:skillscorr}
\end{figure*}


\begin{table*}[tbhp]
\centering
\caption{Important skills and important degrees for different job roles. Quantum skills are in bold text. Importance is defined by consensus, where an important skill/degree has more than 50\% of the respondents indicate it was a primary skill/degree for that job.}
\resizebox{\textwidth}{!}{%
\begin{tabular}{|l|l|l|}
\hline
Job & Important Degree & Important Skills \\ \hline\hline
\makecell[tl]{Applications/Solutions Architect} & \makecell[l]{Masters - Computer Science or Engineering} & \makecell[l]{AI/ML Algorithm Development\\ Applications Design\\ Computer Science \& Engineering\\ {\textbf{Quantum Algorithm Development}}}\\ \hline
\makecell[l]{Circuit Designer} & \makecell[l]{Masters - Electrical Engineering} & \makecell[l]{Analog Circuit Design\\ Digital Circuit Design\\ Electronics\\ Modeling or Simulation}\\ \hline
\makecell[l]{Computational Chemist} & \makecell[l]{PhD - Chemistry/Biochemistry} & \makecell[l]{ {\textbf{Quantum Algorithm Development}}\\ {\textbf{Quantum Science (Chemistry, Physics, etc.)}} \\ Software Development}\\ \hline
\makecell[l]{Control Systems Engineer} & \makecell[l]{PhD - Electrical Engineering \\ Masters - Electrical Engineering \\ Bachelors - Electrical Engineering} & \makecell[l]{Circuit or System Testing\\ Control Theory\\ Noise Measurement \&amp; Analysis}\\ \hline
\makecell[l]{Data Scientist} & \makecell[l]{PhD - Computer Science or Engineering \\ PhD - Mathematics or Statistics \\ Masters - Computer Science or Engineering \\ Masters - Mathematics or Statistics} & \makecell[l]{AI/ML Algorithm Development}\\ \hline
\makecell[l]{DevOps/Database Engineer} & \makecell[l]{Masters - Computer Science or Engineering \\ Bachelors - Computer Science or Engineering} & \makecell[l]{Software Development}\\ \hline
\makecell[l]{Device/Component Engineer} & \makecell[l]{Masters - Electrical Engineering} & N/A\\ \hline
\makecell[l]{Error Correction Scientist} & \makecell[l]{PhD - Physics or Applied Physics \\ PhD - Mathematics or Statistics} & \makecell[l]{ {\textbf{Error Correction}}\\ {\textbf{Quantum Algorithm Development}} \\ {\textbf{Quantum Science (Chemistry, Physics, etc.)}} \\ Theoretical Mathematics or Statistics}\\ \hline
\makecell[l]{Experimental Physicist} & \makecell[l]{PhD - Physics or Applied Physics} & \makecell[l]{ {\textbf{Quantum Photonic/Laser Physics}} \\ {\textbf{Quantum Science (Chemistry, Physics, etc.)}} \\ {\textbf{Quantum Sensor Physics}} }\\ \hline
\makecell[l]{Photonics/Optics Engineer/Scientist} & \makecell[l]{PhD - Physics or Applied Physics \\ PhD - Electrical Engineering} & \makecell[l]{Photonics Circuit Design\\ {\textbf{Quantum Photonic/Laser Physics}} }\\ \hline
\makecell[l]{Quantum Algorithm Developer} & \makecell[l]{PhD - Physics or Applied Physics \\ PhD - Computer Science or Engineering \\ Masters - Computer Science or Engineering} & \makecell[l]{AI/ML Algorithm Development\\ {\textbf{Quantum Algorithm Development}} }\\ \hline
\makecell[l]{Software Programmer} & \makecell[l]{Masters - Computer Science or Engineering \\ Bachelors - Computer Science or Engineering} & \makecell[l]{Software Development}\\ \hline
\makecell[l]{System Assembly/Maintenance Technician} & \makecell[l]{Bachelors - Mechanical Engineering \\ Associate Degree of Vocational School Certificate} & \makecell[l]{Mechanical Assembly}\\ \hline
\makecell[l]{Test/Measurement Engineer} & \makecell[l]{Bachelors - Electrical Engineering} & \makecell[l]{Device Testing \& Characterization}\\ \hline
\makecell[l]{Theoretical Physicist} & \makecell[l]{PhD - Physics or Applied Physics} & \makecell[l]{{\textbf{Quantum Algorithm Development}}\\ {\textbf{Quantum Science (Chemistry, Physics, etc.)}} \\ Theoretical Mathematics or Statistics}\\ \hline
\makecell[l]{the Other role, specfied above} & \makecell[l]{PhD - Physics or Applied Physics \\ PhD - Mathematics or Statistics \\ Masters - Other} & N/A\\ \hline
\makecell[l]{Cryogenics Engineer/Scientist} & N/A & \makecell[l]{Cryogenic Components/Systems\\ Product Development}\\ \hline
\makecell[l]{Product Sales/Marketing} & N/A & \makecell[l]{Business Development\\ Product Development\\ Sales \& Marketing}\\ \hline
\makecell[l]{System Architect/Designer} & N/A & \makecell[l]{Systems Architecture}\\ \hline
\makecell[l]{Technical Sales/Marketing} & N/A & \makecell[l]{Sales \& Marketing}\\ \hline
\hline
\end{tabular}
}
\label{tab:jobskilldegree}
\end{table*}

\section*{Which Skills are Important?}

Building a quantum ready workforce necessitates providing appropriate skills at different educational levels. With this information, students can know how to obtain these skills, colleges and universities can create courses or internships to provide them, and industry can evaluate the usefulness of courses during their hiring process. Crucially, stakeholders need to know from data-driven measures which, and to what extent, quantum skills, in contrast to more traditional skills,  are actually needed by the quantum industry. The answer to this question underpins the value of any quantum course being created by the nation's universities, and guides which courses students should take. 

We define a `quantum skill' as one that is specific to the quantum industry and includes quantum knowledge. We plot the frequency of important quantum and important non-quantum skills in Fig.~\ref{fig:importantskills}, and list them in Tab.~\ref{tab:jobskilldegree}.  

Regardless of the stakeholder, there are important takeaways from Fig.~\ref{fig:importantskills}. First, quantum skills are {\it{not}} needed for all jobs, and non-quantum skills remain important for all jobs. Further, quantum skills tend to cluster by type and into specific job roles. By type, Quantum Algorithm Development (5 quantum skills) and Quantum Science (4 quantum skills) account for 9 out of the 13 occurrences of the important quantum skills. By role, unsurprisingly, it is the more quantum-specific roles that require more quantum skills, e.g., see the error correction scientist role in Tab.~\ref{tab:jobskilldegree}. These data also indicate that many jobs in the quantum industry do not require any important quantum skills. As such, for these jobs, the quantum industry can obtain personnel from the current non-quantum workforce, and people in this workforce should be encouraged to apply to the quantum industry. 

Second, the number of important skills is job dependent. Notably, some jobs require only a single important skill, indicating that a deep knowledge is required. Other jobs however, require many different important skills, indicating that a broad skill set is needed. For the broader skill set, a deep expertise in all skills may not be necessary, and a upskilling course may be sufficient to acquire such skills, rather than a full degree. 

Finally, in Fig.~\ref{fig:skillscorr}, we plot the (Pearson) correlation between skills by averaging the skills over jobs. There are three blocks that are highly correlated among themselves, and which are mostly anti-correlated amongst the others. These skill blocks can be classified as hardware, business, and software. Consequently, these blocks are distinct classes of skills that do not necessitate significant overlap, and personnel should focus only on hardware, business, or software. Moreover, skills within a block tend to be correlated. Universities can use this information to offer the cluster of skills highlighted here, which will in turn enable a robust quantum workforce. 

Our methodology does not consider professional skills (e.g., working on interdisciplinary teams), which are increasingly emphasized by hiring managers in all technology sectors, but not often in posted job ads.  It is likely such skills are  needed here, and therefore should also be included in courses/programs and gained by students prior to entering the workforce \cite{Borner12630,leak2018}.

\begin{figure*}[tbhp]
\centering
\includegraphics[width=.9\textwidth]{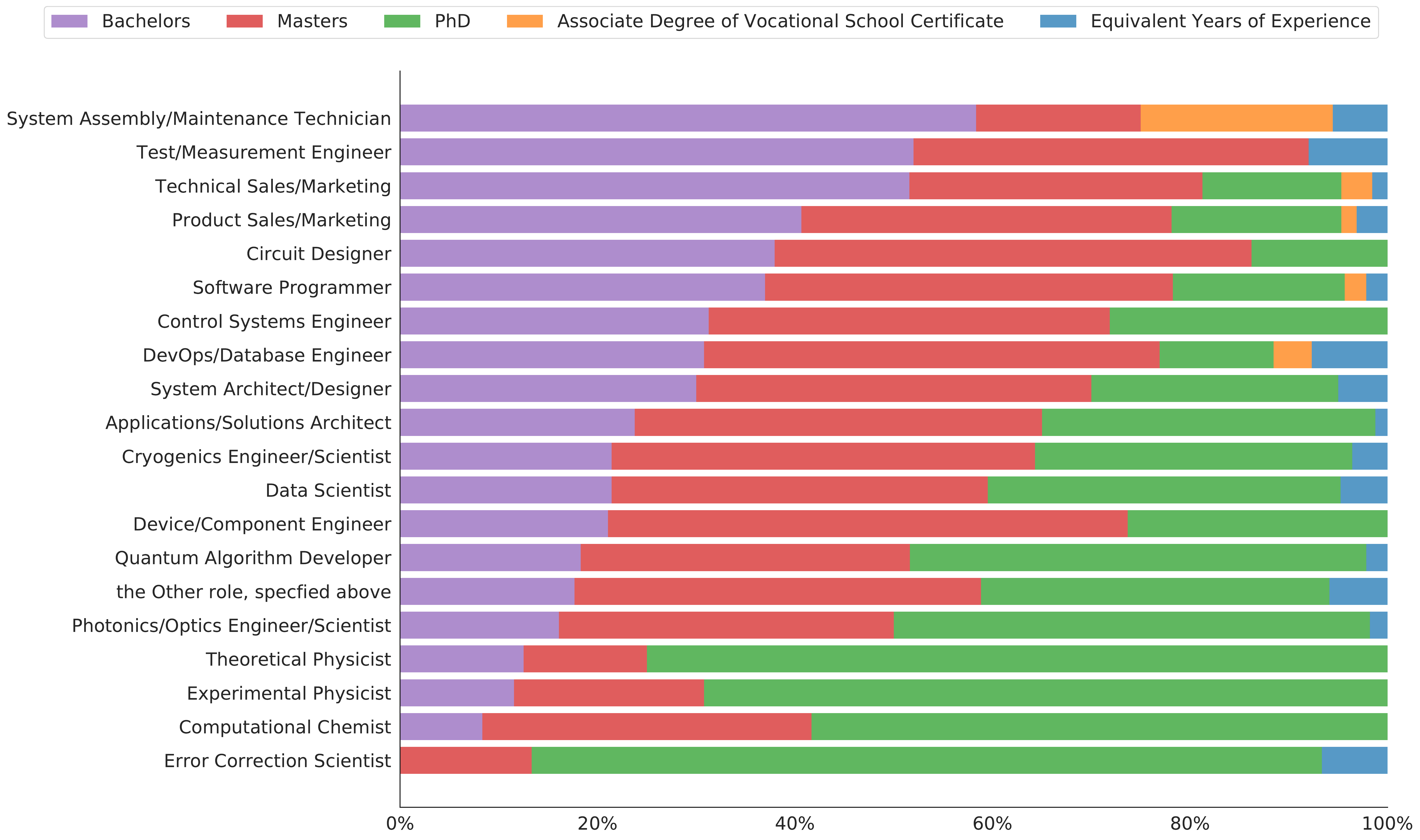}
\caption{The distribution of degrees needed for different job roles in the quantum industry. }
\label{fig:degreesperjob}
\end{figure*}

\section*{Which Degrees are Important?}

In Fig.~\ref{fig:degreesperjob}, we show the distribution of responses regarding the preferred degree for a particular job. This figure provides a valuable insight for universities who want to develop quantum degree programs. For example, if universities are not offering a similar distribution of degrees, they can grow (or contract) offerings to meet market demand. The distributions of Masters degrees follows an intuitive picture, where Masters are needed roughly at the same level for all jobs, except for highly specialised jobs. Jobs requiring a PhD are the quantum specific roles that require important quantum skills, e.g., the error correction scientist. If a PhD is needed for the highly specific quantum jobs, then this places doubt on how many new highly-focused quantum Masters programs are needed for students wanting to enter the quantum workforce. 

The important degrees for each job are given in Tab.~\ref{tab:jobskilldegree}. It should be noted that there is considerably more consensus among which skills are important as compared to which degrees are important. This is understandable, as a degree may or may not provide skills, but it is ultimately the skills and on-the-job experience which are needed. Universities could partner with industries to provide on-the-job experience through internships, thereby making their degrees more valuable. 

As can be observed in Tab.~\ref{tab:jobskilldegree}, Masters and PhDs are frequently considered more important than a Bachelors degree. While members of the workforce with a Bachelors degree have a part to play, particularly in jobs focusing on assembly of hardware, Masters and PhDs are more common. Further, technicians with an associates degree can be included in the quantum workforce, and are likely to be needed increasingly in the future as more companies begin to produce more products. Interestingly, there are certain jobs where there is no consensus on which degree is important; these are shown as N/A  in Tab.~\ref{tab:jobskilldegree}. For these roles however, the important skills are known, which indicates that current degrees may not be providing these skills. Again, this is a useful insight if universities wish to grow to fill a market gap. 

The business roles deserve a notable mention, since they may be overlooked in the quantum workforce. First, business roles make up an appreciable fraction of the job types sought after, as seen in Fig.~\ref{fig:commonjob}. Second, as seen in Tab.~\ref{tab:jobskilldegree}, there are no important degrees for the business jobs, but there are important skills. 

Finally, for jobs which have only important non-quantum skills, a single quantum class in a non-quantum degree program could provide the broad quantum knowledge needed to excel in those quantum jobs.

\section*{Conclusions and Recommendations}

As quantum information science and technology advances, more commercial applications will become possible, and thus precipitate the need for an increasing quantum workforce. To train this new workforce, we must understand what the quantum industry needs in terms of skills and knowledge for different job roles. This information can help inform the development of individual courses, tracks, and degree programs at universities, as well as help students who want to tailor their education to be competitive for quantum-related jobs. Additionally, this information will help those already employed in any sector to know how to enhance their skills and knowledge so as to be able to shift into new quantum-related jobs. 

To inform choices of educators, students, and employees, we report on data from a survey of the quantum industry. First, the analysis of these data showed that there are many different types of quantum-related jobs available now, and that the broad distribution of job types is not predicted to narrow in the near future. This alleviates concerns that the nature of job roles will change dramatically, and the required educational tracks with them. Second, we see that the number and type of important skills and knowledge varies significantly  between job roles. Moreover, many of these important skills are non-quantum skills. This result is crucial for all stakeholders to note, because becoming very specialized in quantum-only skills may not be beneficial for a majority of the workforce.  Third, we show that the degree level required for various quantum-related jobs depends strongly on the job role, where a few jobs essentially require a Ph.D., whereas a majority of roles have no strong preference between a bachelors, masters, or PhD. Currently, there are very few jobs where an associate's degree is valued, but that will likely change in the future as the community moves more into a production phase. Finally, there is also more consensus about which skills are important than about which degrees are important. 

Based on these results, we suggest the follow recommendations:

\begin{itemize}
    \item With so few job roles requiring many quantum skills, educators developing new quantum master's programs should consider the balance between quantum-specific courses and more general STEM courses. 
    \item To prepare students for jobs where there are few or no important quantum skills, universities should consider developing one or two broad quantum courses for this population. Similarly, the quantum industry should be able to acquire appreciable workforce from these quantum-aware graduates. 
    \item Based on the correlations between skills, there is a subset of jobs related to business. These jobs have not been a part of the conversations for quantum workforce development. We recommend that universities begin to engage with leaders in business education on their campuses to prepare students from business majors for roles in the quantum industry. 
\end{itemize}

Finally, we must caution educators and students that the new wave of QIST industry is still young and there are many unknowns about its future \cite{sevilla2020}. Although there is significant promise, there may also be considerable unsubstantiated enthusiasm for how rapidly the science and technology will evolve. Therefore, as new courses and degree programs are developed, we suggest that they prepare students to enter not just the quantum industry, but also more traditional career paths. This would also lead to a larger and healthier STEM workforce overall, which is critical for our economy and nation going forward.

\section*{Acknowledgements}We thank Jon Candelaria for development of the survey questions. This manuscript has been authored by Fermi Research Alliance, LLC under Contract No. DE-AC02-07CH11359 with the U. S. Department of Energy, Office of Science, Office of High Energy Physics. H.J.L acknowledges support from NSF QLCI OMA–2016244. P.M.V. acknowledges support from NSF DMR-1847782.

\bibliography{pnas-sample}

\appendix

\begin{figure*}
\centering
\includegraphics[width=\textwidth]{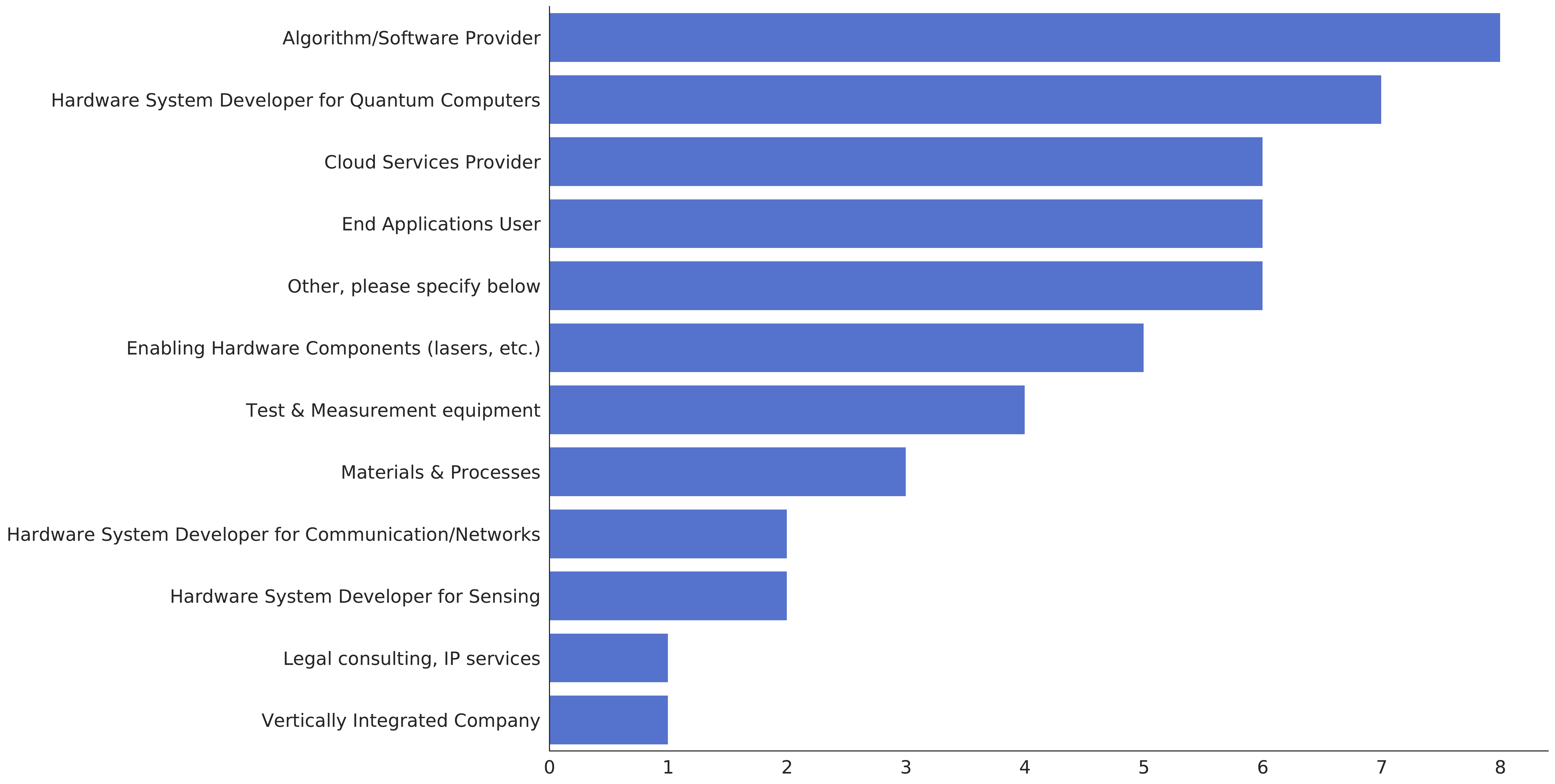}
\caption{Classifications of the 27 companies that self-identified when responding to the survey. The y-axis is the nature of companies, and the x-axis is the number of companies.}
\end{figure*}

\section{Participation Bias Analysis}

The survey was sent to 131 companies, of which 57 responded, and 27 self-identified. However, it is possible that the 57 who responded are not a representative sample of the 131 companies that the survey was sent to, which could introduce  a participation bias. As an example, it would be beneficial to know whether there are more hardware companies in the 57 respondents, when compared to the 131 companies that the survey was sent to. 

To quantify the participation bias in the survey results, we had several authors classify the 131 companies using information on their websites and calculated an inter rater reliability (IRR) between a classification performed by certain authors versus how the 27 self-identified companies classified themselves. If this procedure was successful, it would then be possible to quantify the participation bias. 
In the survey, there are 12 categories that respondents could choose from in order to indicate the nature of their company (see the possible responses to question 1 in the survey in Sec.~\ref{sup:survey}). One of the authors (DAG) classified the entire set of 131 companies that the survey was sent to. Another author (CH) took this classification and compared it against the list of 27 companies that self-identified. Comparing the classification by DAG against the self-reported respondents produced a Cohen's Kappa of 0.33, which is considered only fair agreement. 

As an attempt to improve the value of Cohen's Kappa, a second round of coding was undertaken. In this instance, two other authors (DG and PV), which have different expertise, joined, and all three produced a new  classification through a collaborative coding process. The new author classification was again compared to the self-identified survey respondents values, which increased the value of Cohen's Kappa to 0.46, which is considered only moderate agreement. Both values of Cohen's Kappa are too small to accurately give a reliable estimate of the participation bias, and so we do not report any quantitative measure of this possible bias.

\begin{table*}\centering
\begin{tabular}{l|r|r|r|r|r|r|r|r|r|r|r|r|r|}
Job Role $\downarrow$ \char92{} Company Nature $\rightarrow$
	& \rotatebox{90}{Algorithm/Software Provider} 
	& \rotatebox{90}{Cloud Services Provider}  
	& \rotatebox{90}{Enabling Hardware Components (lasers, etc.)}  
	& \rotatebox{90}{End Applications User} & \rotatebox{90}{Hardware System Developer for Communication/Networks}  
	& \rotatebox{90}{Hardware System Developer for Communication Networks} & \rotatebox{90}{Hardware System Developer for Sensing}  
	& \rotatebox{90}{Legal Consulting, IP Services} & \rotatebox{90}{Materials and Processes}  
	& \rotatebox{90}{Other, please specify below} & \rotatebox{90}{Test and Measurement Equipment}  
	& \rotatebox{90}{Vertically Integrated Company} 
	& \rotatebox{90}{Total}\\ \hline
Applications/Solutions Architect       & 12 &  5 &  2 &  8 &  2 &  3 &  1 &  0 &  0 &  4 &  1 &  3 &  41 \\
Circuit Designer                       &  2 &  1 &  5 &  1 &  3 &  3 &  1 &  0 &  1 &  1 &  4 &  1 &  23 \\
Computational Chemist                  &  2 &  1 &  0 &  1 &  0 &  1 &  0 &  0 &  0 &  0 &  0 &  0 &   5 \\
Control Systems Engineer               &  1 &  3 &  0 &  1 &  1 &  4 &  2 &  0 &  2 &  0 &  1 &  1 &  16 \\
Cryogenics Engineer/Scientist          &  1 &  2 &  5 &  1 &  0 &  4 &  1 &  0 &  3 &  1 &  3 &  1 &  22 \\
Data Scientist                         &  9 &  2 &  0 &  7 &  0 &  1 &  0 &  0 &  0 &  3 &  0 &  1 &  23 \\
Device/Component Engineer              &  2 &  1 &  3 &  1 &  3 &  4 &  1 &  0 &  3 &  0 &  2 &  1 &  21 \\
DevOps/Database Engineer               &  9 &  3 &  1 &  5 &  1 &  2 &  0 &  0 &  0 &  2 &  0 &  0 &  23 \\
Error Correction Scientist             &  1 &  3 &  0 &  1 &  0 &  3 &  0 &  0 &  1 &  0 &  0 &  1 &  10 \\
Experimental Physicist                 &  4 &  6 &  4 &  3 &  3 &  8 &  5 &  0 &  3 &  2 &  2 &  1 &  41 \\
Photonics/Optics Engineer/Scientist    &  2 &  2 &  6 &  1 &  4 &  1 &  5 &  1 &  0 &  1 &  3 &  0 &  30 \\
Product Sales/Marketing                &  7 &  3 &  4 &  2 &  2 &  3 &  0 &  0 &  0 &  3 &  4 &  2 &  30 \\
Quantum Algorithm Developer            & 14 &  6 &  0 & 11 &  1 &  5 &  2 &  0 &  0 &  5 &  0 &  3 &  47 \\
Software Developer                     & 11 &  6 &  3 &  6 &  1 &  5 &  2 &  0 &  0 &  5 &  0 &  3 &  38 \\
System Architect/Designer              &  3 &  1 &  1 &  2 &  2 &  2 &  1 &  0 &  1 &  2 &  0 &  2 &  16 \\
System Assembly/Maintenance Technician &  1 &  2 &  3 &  0 &  1 &  4 &  2 &  0 &  1 &  1 &  4 &  0 &  19 \\
Technical Support/Marketing            &  2 &  2 &  3 &  0 &  1 &  5 &  1 &  0 &  3 &  3 &  3 &  1 &  24 \\
Test/Measurement Engineer              &  0 &  1 &  5 &  1 &  2 &  2 &  2 &  0 &  2 &  1 &  5 &  1 &  22 \\
Theoretical Physicist                  &  2 &  3 &  0 &  1 &  0 &  3 &  0 &  0 &  1 &  0 &  0 &  1 &  11 \\
Other                                  &  1 &  0 &  0 &  1 &  0 &  0 &  0 &  1 &  1 &  2 &  0 &  0 &   6 \\ \hline 
Total                                  & 86 & 53 & 45 & 54 & 27 & 66 & 24 &  2 & 23 & 33 & 35 & 20 & 468 \\
\end{tabular}
\caption{Survey respondents were asked to self-classify the nature of their company (x-axis), and separately to designate which job roles (y-axis) they were seeking to hire within the next two years. The table values give the raw counts from adding the responses from all who replied to both these questions. Note that due to the survey structure, the job role and the company nature are not linked one-to-one. Instead, respondents can reply up to three nature types, and independently any number of job types. As such, these raw numbers must be understood with that caveat. }

\end{table*}

\begin{table*}\centering
\begin{tabular}{l|r|r|r|r|r|r|r|r|r|r|r|r|r|}
Job Role $\downarrow$\char92{} Company Nature $\rightarrow$
	& \rotatebox{90}{Algorithm/Software Provider} 
	& \rotatebox{90}{Cloud Services Provider}  
	& \rotatebox{90}{Enabling Hardware Components (lasers, etc.)}  
	& \rotatebox{90}{End Applications User} & \rotatebox{90}{Hardware System Developer for Communication/Networks}  
	& \rotatebox{90}{Hardware System Developer for Communication Networks} & \rotatebox{90}{Hardware System Developer for Sensing}  
	& \rotatebox{90}{Legal Consulting, IP Services} & \rotatebox{90}{Materials and Processes}  
	& \rotatebox{90}{Other, please specify below} & \rotatebox{90}{Test and Measurement Equipment}  
	& \rotatebox{90}{Vertically Integrated Company} 
	& \rotatebox{90}{Total}\\ \hline
Applications/Solutions Architect       &  9 &  4 &  1 &  9 &  1 &  3 &  1 &  0 &  0 &  5 &  1 &  4 &  38 \\
Circuit Designer                       &  2 &  2 &  3 &  1 &  2 &  5 &  2 &  0 &  1 &  3 &  4 &  1 &  26 \\
Computational Chemist                  &  3 &  2 &  0 &  2 &  0 &  1 &  0 &  0 &  0 &  0 &  0 &  0 &   8 \\
Control Systems Engineer               &  1 &  2 &  1 &  0 &  1 &  3 &  1 &  0 &  2 &  0 &  0 &  0 &  11 \\
Cryogenics Engineer/Scientist          &  2 &  3 &  3 &  1 &  0 &  4 &  1 &  0 &  1 &  1 &  3 &  1 &  20 \\
Data Scientist                         &  9 &  2 &  1 &  5 &  1 &  1 &  0 &  0 &  0 &  4 &  0 &  0 &  23 \\
Device/Component Engineer              &  2 &  2 &  3 &  1 &  1 &  5 &  2 &  0 &  2 &  2 &  4 &  1 &  25 \\
DevOps/Database Engineer               &  7 &  3 &  0 &  5 &  0 &  2 &  0 &  0 &  0 &  3 &  0 &  0 &  20 \\
Error Correction Scientist             &  2 &  3 &  0 &  2 &  0 &  3 &  0 &  0 &  1 &  0 &  0 &  1 &  12 \\
Experimental Physicist                 &  5 &  5 &  6 &  3 &  3 &  8 &  7 &  0 &  3 &  2 &  4 &  1 &  47 \\
Photonics/Optics Engineer/Scientist    &  0 &  1 &  5 &  1 &  4 &  5 &  5 &  0 &  0 &  2 &  4 &  0 &  27 \\
Product Sales/Marketing                &  6 &  3 &  8 &  2 &  1 &  5 &  1 &  0 &  1 &  3 &  5 &  1 &  36 \\
Quantum Algorithm Developer            & 13 &  5 &  0 & 10 &  0 &  4 &  2 &  0 &  0 &  5 &  0 &  2 &  41 \\
Software Developer                     & 11 &  5 &  2 &  7 &  3 &  4 &  0 &  0 &  1 &  3 &  1 &  1 &  38 \\
System Architect/Designer              &  4 &  2 &  1 &  4 &  2 &  1 &  3 &  0 &  1 &  4 &  0 &  1 &  23 \\
System Assembly/Maintenance Technician &  2 &  2 &  3 &  0 &  0 &  4 &  2 &  0 &  0 &  1 &  3 &  1 &  18 \\
Technical Support/Marketing            &  5 &  5 &  1 &  2 &  3 &  5 &  1 &  0 &  3 &  3 &  3 &  2 &  33 \\
Test/Measurement Engineer              &  2 &  1 &  7 &  1 &  4 &  1 &  1 &  0 &  1 &  1 &  5 &  1 &  25 \\
Theoretical Physicist                  &  5 &  4 &  0 &  2 &  0 &  5 &  1 &  0 &  0 &  0 &  0 &  1 &  18 \\
Other                                  &  1 &  0 &  0 &  0 &  0 &  0 &  0 &  1 &  0 &  0 &  0 &  0 &   4 \\ \hline 
Total                                  & 91 & 56 & 45 & 58 & 26 & 69 & 30 &  1 & 17 & 44 & 37 & 19 & 493 \\ \hline 
\end{tabular}
\caption{Survey respondents were asked to self-classify the nature of their company (x-axis), and separately to designate which job roles (y-axis) they were seeking to hire within the next 3-5 years. The table values give the raw counts from adding the responses from all who replied to both these questions. Note that due to the survey structure, the job role and the company nature are not linked one-to-one. Instead, respondents can reply up to three nature types, and independently any number of job types. As such, these raw numbers must be understood with that caveat. }

\end{table*}

\FloatBarrier
\section{Survey}
\label{sup:survey} 
The following  contains the contents of the survey we administered. We present general results from the collected survey data in the main text.

\clearpage
\includegraphics[page=1, width=1.0\textwidth]{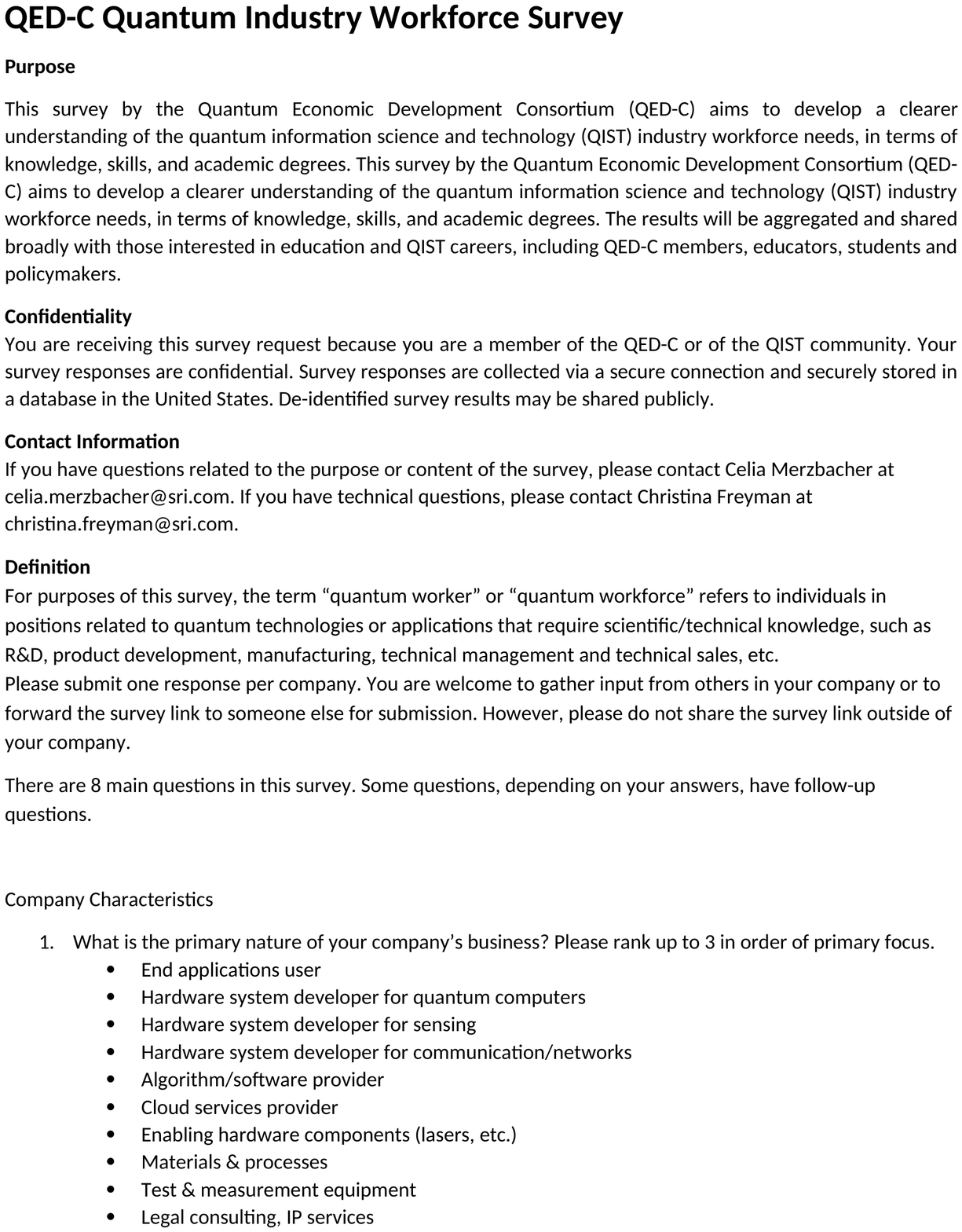}
\clearpage
\includegraphics[page=2, width=1.0\textwidth]{survey_final.pdf}
\clearpage
\includegraphics[page=3, width=1.0\textwidth]{survey_final.pdf}
\clearpage
\includegraphics[page=4, width=1.0\textwidth]{survey_final.pdf}
\clearpage
\includegraphics[page=5, width=1.0\textwidth]{survey_final.pdf}

\end{document}